\documentclass{article} 
\usepackage{iclr2025_conference,times}

\usepackage{amsmath,amsfonts,bm}

\def\eqref#1{equation~\ref{#1}}

\def\1{\bm{1}}

\DeclareMathAlphabet{\mathsfit}{\encodingdefault}{\sfdefault}{m}{sl}
\SetMathAlphabet{\mathsfit}{bold}{\encodingdefault}{\sfdefault}{bx}{n}

\usepackage{hyperref}
\usepackage{url}
\usepackage{graphicx}
\usepackage{grffile}
\usepackage{tabularx}
\usepackage{booktabs}

\title{The human visual system can inspire \\ new interaction paradigms for LLMs}

\author{Diana Robinson \& Neil Lawrence \\
Department of Computer Science and Technology\\
University of Cambridge\\
Cambridge, UK \\
\texttt{\{dmpr3, ndl21\}@cam.ac.uk} \\
}

\iclrfinalcopy 
\begin{document}

\maketitle

\begin{abstract}
The dominant metaphor of LLMs-as-minds leads to misleading conceptions of machine agency and is limited in its ability to help both users and developers build the right degree of trust and understanding for outputs from LLMs. It makes it harder to disentangle hallucinations from useful model interactions. This position paper argues that there are fundamental similarities between visual perception and the way LLMs process and present language. These similarities inspire a metaphor for LLMs which could open new avenues for research into interaction paradigms and shared representations. Our visual system metaphor introduces possibilities for addressing these challenges by understanding the information landscape assimilated by LLMs. 
In this paper we motivate our proposal, introduce the interrelating theories from the fields that inspired this view and discuss research directions that stem from this abstraction.
\end{abstract}

\section{Introduction}
\label{introduction}

Human visual perception gives us the illusion that we perceive an entire visual landscape when in reality our eyes are only capable of detailed resolution in a small part of the scene. This illusion is enabled by ocular fixations that move rapidly across the visual space in movements known as saccades \citep{Snowden2012-di, Mackay-behind91, ORegan-whyred19}. Similarly, large language models (LLMs) are able to generate highly coherent natural language responses in ways that seem to suggest that they understand a wider context. Like the eye they interpolate over areas where they lack information, leading to what has been termed hallucinations. In our ocular system the equivalent interpolations can manifest as visual illusions. Through these interpolations, both systems can provide an impression of a fuller representation, visual or conceptual, than is actually present. We argue that this similarity between human visual perception and the way LLMs process data could open promising new avenues for research into interaction paradigms by leveraging shared representations, information exploration, and defining informational affordances to evaluate the success of interactions. 


With the explosion over the last few years in the use of LLMs \citep{Vaswani2017-wa, Devlin2018-zp, Enis2024-pq, OpenAI2023-ig} comes new urgency to understand just how we should conceptualise LLMs as we design systems that exploit their capabilities. We will explore the relationship between the way we perceive the visual world and the workings of LLMs and see how it informs our understanding. Our work contributes to the need for research into humans' perceptual adaptation to AI in order to critique and collaborate with AI systems \citep{Shen2024-rg}. Developing our metaphor and the research directions below can lead to new ways for humans to think critically about LLMs, to develop the right degree of trust in LLM outputs and calibrate their mental models of how LLMs operate \citep{Shen2024-rg}.

Several authors have provided metaphors to develop our understanding of LLMs \citep{Mitchell2024-bj, Trott2024-jg}. They range from ``stochastic parrots'' \citep{Bender2021-oj} to role players \citep{Shanahan2023-wt} to ``blurry JPEGs of the web'' \citep{Chiang2023-mz}. The use of metaphor can be a helpful shorthand to emphasise particular qualities of LLMs but in providing a shortcut to understanding, metaphors necessarily elide other characteristics \citep{Trott2024-jg}. We discuss the limitations to our own choice of metaphor in Section \ref{alternative-views}. The metaphors we choose matter: they influence levels of trust in LLMs, the understanding we ascribe to them, legal arguments governing their use, and which experiments they lead us to perform in seeking scientific understanding of them \citep{Mitchell2024-bj}. For example, the LLMs-as-minds metaphor leads to scientific avenues of experimentation such as standardised tests, such as IQ, designed for humans being applied to LLMs \citep{Mitchell2024-bj}. Conversely, viewing the LLM as a blurry JPEG might trigger us to think more in terms of compression and accuracy of reconstruction. Metaphor can also be a guide to designing interfaces for interacting with information systems \citep{Neale1997-jp}. For example, the desktop metaphor, developed decades ago at Xerox Parc \citep{Voida2008-jz} has been an enduring one that uses the physical office space to give us a shorthand to navigate the digital information space on our computers.  


In this paper we explore whether we can use the shared similarity between our assimilation of the visual world through our visual perception and the representations that LLMs build up from studying language to inform our understanding and use of LLMs with better insight into their limitations. The analogy to visual perception also allows us to build on a foundation of research in visual perception to inspire new research directions that allow us to envisage new interaction paradigms for LLMs. 

In her lectures on trust, Baroness Onora O'Neill argues that trust between humans is developed through the elicitation and evaluation of salient pieces of information rather than an onslaught of every possible piece of information, arguing that transparency is not the same as trust \citep{ONeill2002-jg}. Determining the right degree of trust is essential to understanding how and in what ways to integrate LLMs into tasks that would previously have been the domain of humans. LLMs can make it seem like there is shared understanding which can cause them to err in potentially catastrophic ways if deployed in high stakes contexts such as healthcare or law.  By better understanding the ways in which LLMs assimilate and produce information, we believe we can more effectively interrogate salient information they are relying on to produce their responses, leading to developing the right degree of trust. 


Classically, interactions between humans and machines are explored through \emph{interfaces}. For contextualising our metaphor, we prefer the term interaction paradigm to capture the richer nature of the human-machine affordances that these models enable. We can split the interaction paradigm into three parts. Firstly, a shared conceptualisation between the LLM and a digital artefact of interest (e.g. a collection of academic papers, a software codebase, a mathematical proof). Secondly, an interface between the LLM's realisation of that artefact and the human being (see, for example, the notion of the human analogue machine \citep{Lawrence2024-qm} in section \ref{shared-representations}). Thirdly, the interaction paradigm may also develop better ``cultural perception'' of these models, enabling improved training efficiency, better understanding of limitations of current architectures, and progress toward resolving open challenges like hallucinations.

\section{Shared representations}
\label{shared-representations}

Early research in vision theorised that as we look at our environment, we develop an internal representation of it, like a small pictorial depiction of the external world inside our heads \citep{Marr2010-gv, Lorenz1978-ku}. Under this framework, it is assumed we are able to make sense of the world visually by manipulating the representation. Our thoughts are enabled through exploratory actions in an imaginary space with a ``neural `model' of outer reality'' \citep{Lorenz1978-ku}. However, the limitations to this view revealed themselves in different visual paradoxes that required more convoluted mechanisms to explain. For example, why wasn't there a seam in the visual field, if we were adding it together with information from each eye? \citep{ORegan-whyred19} And if the eyes move in saccades, why is there not a blur in the image each time the eyes move in this way? \citep{ORegan-whyred19} 

\subsection{External visual memory}

A new line of thinking, revolutionised this conception of visual perception by instead suggesting that there is in fact no representation of the world inside our heads but the world around us functions as ``external visual memory'' \citep{Mackay-behind91, ORegan-whyred19, ORegan-outside92, Gibson-ecological79, Merleau-Ponty2013-ts}.  Neuroscientist and information-theorist Donald MacKay suggested we should think about the eye as ``a giant hand that samples the outside world'' \citep{ORegan-whyred19} an idea also reflected in the thinking of psychologist James \citet{Gibson-ecological79} and philosopher Maurice \citet{Merleau-Ponty2013-ts}. This led to breakthroughs in conceptualising both representation and perception that led to useful experiments to verify the concepts (see for example \citep{ORegan-outside92, ORegan-whyred19}). And it is these concepts that we argue are a basis for drawing useful parallels with LLMs. 
 
MacKay's proposition was that visual perception is not an ``input-throughput-output system'' but instead the brain is organised as a conditional probability network, processing visual inputs in a state of ``conditional readiness for action'' toward particular goals \citep{Mackay-behind91}. By taking this view, perception was then characterised as updating the state of conditional readiness based on the inward information flow of sensory inputs \citep{Mackay-behind91}. This made attention a central focus of vision with the region of attention under active matching being the focus of perception \citep{Mackay-behind91}. This importantly explained several visual illusions such as ``change blindness'', where large changes in visual field are not perceived when there is a brief visual disruption \citep{O-Regan1999-sa}.

\subsection{Grounding exploration in shared representations}

\begin{figure}[h]
\begin{center}
\includegraphics[width=0.8\linewidth]{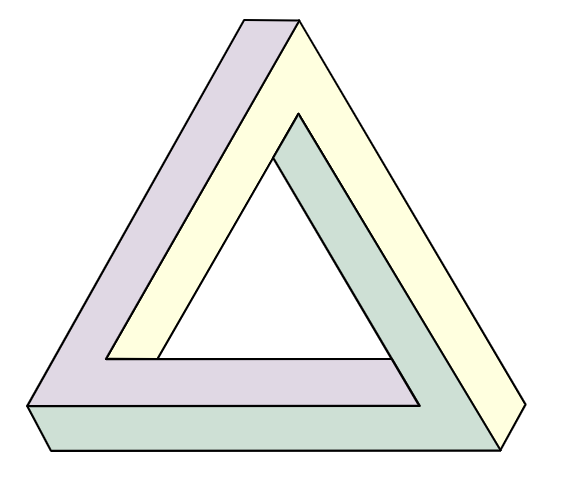}
\end{center}
\caption{The Penrose Triangle. While the image is locally consistent, it is globally inconsistent. In our metaphor a similar inconsistency could occur in an LLM where a conversation is consistent in individual exchanges, but inconsistent in its broader structure. Such a conversation would be poorly grounded.}
\label{fig:penrose-triangle}
\end{figure}

These researchers made a conceptual connection between physical exploration and visual exploration and we now argue for a further connection to exploration of the abstract information space. An important difference between the physical world and the space of knowledge of LLMs, written information, is the absence of a shared representation. A key aspect of this difference can be the lack of ground truth. In the physical world, even if we are tricked by visual illusions, these can be revealed to us to reverse the illusion with reference to a shared external world. However, unless the model is grounded in a particular digital artefact there is no immediate analogue in the information space. This problem of lack of shared representations is sometimes referred to as knowledge grounding \citep{Huang2024-nb}. See Figure \ref{fig:penrose-triangle} for a visual illusion characterising a poorly grounded shape in the form of a Penrose Triangle. Huang et al. highlight the importance of temporal shifts in information where LLMs do not incorporate new information. 

\subsection{Cultural biases}

For LLMs these challenges can manifest more insidiously. For example, LLMs are trained on our written language, code and images, but not all human knowledge is recorded in this way. And even within written knowledge, there are culturally embedded differences in language such as the ways in which the usage of words evolve over time and their meanings evolve with use \citep{Wierzbicka2006-uu}. As an illustrative example, \citet{Wierzbicka2006-uu} studied the evolution of the usage of the word ``fair'' in English and mapped its conceptual shift to mean procedural fairness, while this word and associated implications are missing from other languages. So the language used to train LLMs can markedly influence the culturally loaded meanings embedded within them. The over-representation of English and Chinese language models means that there are blind spots in building up a culturally accurate representation of knowledge. Additionally, psychologists have characterised the dominant understanding in the west as WEIRD \citep{Henrich2010-is} suggesting that ideas originating in the west are psychologically peculiar. This psychological peculiarity is likely mirrored in LLMs and is problematic because it may undermine the broader diversity of human thinking and the agency of individuals from different cultures. 

\subsection{Human Analogue Machines}

The fluency of LLM responses can elide the very different processes by which humans reason and LLMs process data \citep{Lawrence2024-qm}. Lawrence proposes the concept of Human-Analogue Machines (HAMs) to refer to the reflection of ourselves that LLMs learn from training on our written data. The idea is that the LLM has learnt a representation space that is somehow analogous to the feature landscape we use to generate and understand language (see also \citep{Vallor-aimirror24}). So while the LLM does not feel and perceive as a human does, it provides a representation of cultural ideas that allows it to bridge between the machine's enormous communication bandwidth (billions of bits per minute) and the human's (thousands of bits per minute) (see also \citep{Zheng2025-jz}). This means that LLMs provide a new interface between our thinking and the digital representations that they can assimilate. The LLM is bridging a significant information asymmetry. Similarly,the eye bridges between the information available in the visual space and the limited information we can absorb through sensors on our retina and the bandwidth of the optic nerve. 

We cannot fathom how we would `think' if we had direct access to the vast bandwidth that LLMs have. Further, while an LLMs' human context is developed in an analogous feature space, an LLM does not have a social stake in the way a human does. It can emulate human ideas and cultural understandings but it does not originate them in the same manner humans do. For this reason we would like humans to maintain their agency when interacting with an LLM. 

The visual system offers a useful metaphor because in visual perception we cannot take in the entire visual scene but we perceive through fixations and saccades that are influenced both by features present in the image and steered by the tasks we are looking to perform. This then provides the sense we have of understanding the whole picture and being able to make our decisions accordingly \citep{Snowden2012-di,Triesch-need2003}. This understanding is mediated by the ocular system. 

As LLMs increasingly become an interface between our thinking and the digital representations they mediate between \citep{Lawrence2024-qm, Vallor-aimirror24}, this becomes problematic as we cannot audit the totality of their information processing. It can be difficult to understand to what extent they are misrepresenting the information we provide them and to judge how they are balancing that information against wider human context. This isn't only important in safety critical domains, but in our every day interactions where our agency may be undermined by persuasive descriptions of data that could be based on flawed representations of information or a miscalibrated assimilation of the human context.

\subsection{Case study}

To ground our ideas, we would like to highlight an application of LLMs in mathematics that provides a concrete example of a type of shared representation that allows for anchoring of meaning between multiple people. The Equational Theories project pioneered new ways of collaborating with machine assistance on open mathematical challenges in the area of universal algebra \citep{Bolan2024-dc}. In this context, the mathematical proof provided a stable point of reference so that different team members could work on aspects of the larger problem in tandem. They were able to take a proof and write a blueprint that broke it into steps to formalise separately using software to translate formal proofs into human-readable proofs. This allowed distributed team members the ability to expand and look at hypotheses to prove in the larger context of the overall project. The shared representation enabled collaboration and division of labour while not losing the overall goal. This is a case study of the interplay between a representation and domain users which grounds efforts to solve a problem in shared ground truth and understanding. We will highlight aspects of the above case in more detail in the sections that follow to illustrate the research directions that we argue will be most productive to explore with respect to the analogy of LLMs to visual perception. 


\subsection{Research directions}

So what might a shared representation of the information space traversed by an LLM look like? Again, we look to vision research. Saccade mapping or salience mapping is used in studies of visual perception to develop an understanding of what information our eyes rely on to interpret and process an image, such as another human face \citep{Tatler2006-yk}. Understanding where the eye fixates provides a basis for making testable hypotheses about how visual information is assimilated under different conditions. Similarly, developing an understanding of how the attention mechanism in transformer architectures underlying LLMs focuses across the information space could help to develop an understanding of the depth and breadth of coverage across a set of knowledge.

\subsubsection{A saccade map of the information space}

Concretely, we ask whether it is possible to make a saccade map of abstract concept space that LLMs are leveraging in their responses. What would a map of the concept space look like in terms of the meanings and connections between tokens that LLMs rely on to generate responses? Can we visualise this and reveal misleading or incorrect information and areas where we need to prompt them to assimilate other information to generate correct responses? 

Apart from developing a better way to audit LLM responses, a saccade map of the concept space, illuminating better the workings of the attention mechanism, could yield more efficient training strategies by leveraging potential patterns in attention across words. 


\subsubsection{Building human decision making into algorithmic architectures}

One way to investigate the possibility of developing such a map can be seen in the work of Sanborn and Griffiths in which they develop a method for inserting humans into the acceptance function of a Markov Chain Monte Carlo algorithm in order to elicit their subjective probability judgements \citep{Sanborn2008-gj}. More broadly, they suggest using people as elements of machine learning algorithms to test hypotheses about how humans reason. The converse would also be interesting to investigate, can we develop strategies for doing this which could yield better understanding of the workings of particular algorithms? We suggest a possible research direction to introduce human choice into transformer architectures in similar fashion to the MCMC with people in acceptance function. It would function similarly to reinforcement learning with human feedback (RLHF) \citep{Askell2021-bb, Ouyang2022-mv, OpenAI2022-pg} but instead of humans vetting the final response, they would be assessing information internal to the transformer architecture for defining and refining useful levels of abstraction of concepts. As the transformer looks at a token in context of all the other vectors of meanings, human feedback could steer it to toward an appropriate level of abstraction through concept editing. This line of work would require new visualisations of spaces of concepts at a level where LLM responses can be contextualised and assessed to determine a level of appropriate knowledge.

\section{Exploration and information seeking}
\label{exploration}

We have discussed shared representations as ways of checking whether either the visual image our eye has built up or the LLM responses track reality and the need to do this arising from information asymmetries. But how is this exploration of information in order to prioritise it done? We will start with the vision case. Donald MacKay compares the structure and function of the visual system in the way that it filters information to an 'organisation' to contrast with the information channel view. He compares visual processing in which a large amount of visual information is discarded in selective processing to how companies are structured, such as a sales organisation where individual transactions are happening at the lower levels but the majority of this sort of granular information is filtered out selectively on the way up to the higher levels of management to enable them to make strategic decisions \citep{Mackay-behind91}. This captures from a different perspective how the visual system mediates the image from the environment that is received. But how do we build up this picture of a visual landscape through saccades? We have discussed saccade mapping as a way of seeing where the eye fixates but how is the choice on locations to fixate made?

\subsection{Top-down, bottom-up, and systematic saccades}

Experimental evidence suggests the influences of both bottom-up and top-down approaches to eye movements \citep{Schutt2019-yr}. Bottom-up influences were theorised from anti-saccade experiments where subjects were told not to look at a particular stimuli such as a flashing light but their eyes were nonetheless drawn to it without their conscious control \citep{Mokler1999-wk, Munoz2004-xa}. Top-down gaze fixation has been studied in the context of gaze fixations in response to different task prompts \citep{Einhauser2008-kn}. A third category has been proposed, systematic approaches, which refer to common patterns of gaze fixation such as a preference to focus on the centre of an image or to saccade in some directions more frequently than others \citep{Tatler2008-hp}. Some experimental evidence suggests that saccade patterns show statistical dependencies to prior saccades and this could be a key connection between the attention involved in guiding this type of saccadic behaviour and the attention mechanism of LLMs as they contextualise tokens in a sentence against other nearby tokens probabilistically.  The relative influence and the ways in which these trade-off with each other are an open topic of research \citep{Schutt2019-yr}.

The differentiation in visual perception between bottom-up, top-down or systematic saccades can be a productive line of inquiry in thinking about interactions with LLMs, i.e. are we being task-driven or are we responding and changing based on the responses provided, and is there a particular standard pattern of interaction common to most users that results from a shared ability to read and process written language? We can imagine many cases of task-driven exploration where LLMs are being used for summarisation of documents, to answer particular questions or to gain a broad understanding of a topic. Conversely, sometimes LLMs might be used in an exploratory sense, for example in the context of creative brainstorming. Even in the former case of understanding a topic, this blurs the distinction between top-down and bottom-up in the sense that it is goal-directed but also involves elements of exploration because it takes exploration to assess the right degree of understanding.

In machine learning the trade off between exploration and exploitation is driven by time horizons and levels of uncertainty. This suggests that these trade-offs should also be task dependently driven when exploring information landscapes through conversation. This idea maps nicely onto the top-down and bottom-up influence that visual perception researchers refer to in saccade modelling. 


\subsection{The sensorimotor approach to vision and theory of active inference}

The sensorimotor approach to vision \citep{Gibson-ecological79, ORegan-whyred19, Mackay-behind91} that we have been outlining can be usefully aligned to a proposed theory of active inference (AIF) to model human-computer interaction  \citep{Murray-Smith2024-vr}. In this, Murray-Smith et al. outline three models, a model of the user, their environment, and the interface. Their interface model is defined as of ``an interface that looks in both directions to black box user and system models''. This relates well to Lawrence’s concept of Human-Analogue Machines introduced earlier in which the LLM mediates interactions between humans and digital representations \citep{Lawrence2024-qm}. We are interested in extending and applying this thinking in information spaces of LLMs where AIF is more typically associated with closed loops involving sensors and the physical world.

\subsection{Case study}

In the mathematical collaboration case study, the core structure of the proof coordinates the exploration patterns we described earlier. This structure enabled mathematicians to expand from narrow hypotheses toward the larger goal. The proof's division of labour facilitated explore/exploit behaviour, where collaboration on specific proof elements explored possible solutions that could then be verified in context.

When interacting with LLMs, which lack working memory, the shared representation of the proof serves as a stable reference point. This helps track progress toward both smaller hypotheses and the overall proof goal.




\subsection{Research directions}

Top-down, bottom-up and systematic saccades provide a useful model for thinking about different levels of attention in LLMs and how they might be explored. The type of features picked up by different saccade strategies could relate to different levels of abstraction of concepts in information space and navigating these could provide a way of exploring the ``informational saccade map'' proposed in section \ref{shared-representations}.

There is evidence eye movements behind the saccade are heavy-tailed in their distribution \citep{Tatler2006-yk}. This means that the eye follows a search strategy of small, frequent jumps for detailed exploration, with occasional large jumps to survey the broader landscape and discover new regions of information. This is reminiscent of statistical sampling and Markov chain Monte Carlo algorithms where heavy-tailed Markov chains have theoretically better performance \citep{Jarner-heavytail07}. This suggests that when exploring the information landscape through conversation, heavy tailed movements may also be appropriate.

Computational approaches such as those used by \citet{Schutt2019-yr} have been used in analysing visual saliency. Could similar approaches be used in studying the information seeking exploratory behaviour of human users of LLMs while engaging in both goal-directed or exploratory search of the information landscape mediated by LLMs? Perhaps this could inform the user model within Murray-Smith’s theory of active inference \citep{Murray-Smith2024-vr}. Active inference models could provide a framework for modelling top-down interactions for evaluating particular pieces of the information landscape against user goals. This could give us a way to quantify both the user evaluation and the feedback loop as information is assessed against the goals of the user and then accepted or rejected.



O'Regan describes the role of attention in seeing \citep{ORegan-whyred19} and uses this understanding to design experiments that confirm that the area of focus of attention results in ``change blindness'', the phenomenon that when we are focusing on a given part of a visual scene, we can easily miss large changes in another aspect due to the fact that we are not dedicating attention to it. He introduces the term ``grabbiness'' to describe how particular visual stimuli such as sudden noises or flashes of light grab attention, regardless of prior focus. This type of mechanism could usefully be applied in exploration of LLMs with the ``flashing light'' in this case being distribution shift and other relevant indicators that data is scarce and the exploration has gone wrong in some important way. Out of distribution data is a critical failure mode in modern AI systems \citep{Marcus1998-ca, Marcus2024-yi}. Can we design something like ``grabbiness'' in LLMs that can pull focus to distributional shift, signalling the user to investigate and add in context? This would usefully indicate when the user and the LLM were out of alignment or that their understanding was starting to diverge, so that the human user could intervene with a different level of attention.



\section{Affordances}
\label{affordances}

Affordances is a term coined by psychologist, James Gibson, that characterises the relationship between an animal and its environment by describing the type of information that allows the animal to understand its particular action possibilities (for example, if a surface is nearly horizontal, nearly flat, sufficiently extended and rigid, it is ``stand-on-able'' for humans) \citep{Gibson-ecological79}. Gibson describes the way in which we are able to visually understand affordances based on information within them and our perceptual system. This departs radically from previous views from empiricism theorising that we understand meaning and value of objects based on past experience \citep{Gibson-ecological79}. Gibson argues for the view of a perceptual system made up of organs with different capacities for exploration as a basis for vision versus an information channel approach to vision by which stimulus information is communicated from the environment that is then acted on by the organism internally. 






In another aspect of Gibson's theory of perception, he describes knowledge mediated by descriptions (communication of particular concepts or affordances without the listener directly experiencing them) in which crucially the aspect of reality testing is missing. He differentiates between explicit versus tacit knowledge (which comes from direct experience). What does this distinction mean in the information space and mediated by LLMs? We turn back to the case study to explore a possible mapping from the ideas of physical exploration to the information landscape.



\subsection{Case study}

In the mathematical collaboration case  \cite{Bolan2024-dc}, the role and function of proof is built up through domain understanding of mathematics and using this as an anchor point allows for useful interactions with it. The affordances of the proof for the mathematicians is clear but understanding of these affordances is built up (in Gibson's term ``mediated through description''). The mathematicians need to undergo formal training, and mathematical proofs need to be converted into words and human concepts before exploration.

Mathematical proof is an interesting case in which in some sense the concepts can be ``directly experienced'' in a similar way to perceiving the physical world in that theories can be tested against a structured framework. It is mediated by descriptions but the possibility for exploring and verifying within this abstract space is interesting and seems different in an important sense from Gibson’s description of knowing mediated by descriptions where we cannot understand concepts without directly experiencing them. There could be an extrapolation to make here from the type of exploration that yields affordances in the information space. It is this that we would like to pick up on and experiment with in the research directions outlined below.

\subsection{Research directions}


Affordances could provide a conceptualisation for domain understanding against what is captured in the LLM. The concept of affordances has been much discussed in human-computer interaction starting with Don \citet{Norman-psychology88} and continuing into thinking about user interfaces and possible theories stemming from affordances as construed within social and cultural contexts \citep{Kaptelinin2012-od}. There is a prevailing view that Gibson's theory of affordances needs to be regrounded and defined for interactions that involve these larger contexts \citep{Kaptelinin2012-od}.

In our particular case drawing the analogy between LLM exploration and visual perception, we return to Gibson's original conception to see where it might lead for the information landscape of LLMs. Gibson defines qualities of objects that influence how our perception of them allows us to understand their affordances. One of the examples he uses is of ``landing judgment’’ in training pilots, which is the determination of a particular point at which it is essential to aim that orients the pilot between their motion and the external world while landing a plane \citep{Gibson-ecological79}. \citet{Gibson1955-vi} mathematically defined this particular point in order to measure and experiment with it. 

This concept of affordances and how they might be measured in the above example can inform our thinking about exploration of the information space of LLMs. How might we translate this notion of affordances and possibly quantify it in relation to humans being able to assess the level of concept mapping that LLMs have assimilated in order to understand whether the responses can be usefully integrated into their tasks? Can we mathematically define concepts  in the information space of LLMs, similarly to the way that Gibson defined ``landing judgment'' in the physical space, at a level at which we could recognise and audit them? And could we customise them to the level of abstraction needed?


Returning to Gibson's original conception of affordances and past vision experiments could inform a technical mapping between informational affordances of LLMs and how they can be applied in different user-oriented contexts. We can then take the consequences of this technical and conceptual understanding into the ways in which we place LLMs in domain contexts and use interfaces to communicate underlying information structures.

\section{Alternative Views}
\label{alternative-views}

We have argued for the fundamental similarity between visual perception and the ways in which LLMs process information as the basis for our argument for the three sets of research directions we propose for LLM interaction paradigms. This relies on the similarity we outline in assimilating the physical world or the information landscape being importantly similar. The alternative view is that this similarity does not hold.

One way in which this might not be true is that the compression needed to be done by an LLM to communicate with a human versus the eye processing the visual landscape is of an order of magnitude that they are not importantly similar. We calculate this difference in compression in the following way. The ratio between machine to machine communication vs human to human communication is roughly 300 million. So when using saccades and fixations with an LLM we are attempting to view an object using a bottleneck of 1/300,000,000. We can make a rough calculation to compare this with the information asymmetry the eye is dealing with. The mid-peripheral field of vision of the eye is around 120 degrees. The minimal angle of resolution (MAR) for the eye is estimated at 60 pixels per degree \citep{Strasburger-peripheral11}. For each pixel it's estimated that we can resolve  around 10 million colors in typical lighting conditions our temporal resolution is around 30 frames per second. These numbers allow us to compute the maximum availability of resolvable information in our mid peripheral visual field as $60^2\cdot 60^2\cdot\pi\cdot 24\cdot 30$ bits per second or around 30,000 Mega bits per second. Estimates of the eye's information bandwidth suggest the optic nerves can carry around 10 Megabits of information per second \citep{Koch-eyebrain06}. This implies an information asymmetry factor of only 3000, or 100,000 times lower than our estimate of the asymmetry between the LLM and human. This gap between what the eye needs to do to summarise the ``big picture'' vs what the LLM would need to do may imply that the approaches used by the eye are not appropriate for an LLM.

The information in the visual field also has particular structure, it lives in an evolving three dimensional space and is characterised by the reflective properties of light. The LLM's information landscape is likely to have a much higher intrinsic dimensionality. This might render the notion of a saccade in the space less useful in capturing the information appropriately for the human subject. These differences between a physical space and a less tangible information space might suggest that the two spaces of interaction are sufficiently different that a comparison between interaction behaviours is misleading.

We argue that though the differences outlined above are relevant, the utility of the metaphor will be proven or disproven in the pursuit of the research agenda that we propose above and in the experiments that follow from conceiving of LLM interaction paradigms in this way.


\section{Conclusion}
\label{conclusion}

In this position paper, we have argued that the visual system can inspire new interaction paradigms for LLMs. We discuss the fundamental similarity between the ways in which we build up a visual picture of the world through saccadic eye movements and the ways in which LLMs build up an information context through tokens. We introduce foundational concepts from the sensorimotor approach to visual perception: shared representations, exploration and information seeking, and affordances, and propose key research directions for interaction with LLMs that stem from these. A summary of the key theories of visual perception that we draw upon and their connection to LLM research directions is provided in Table \ref{summary-table} in the Appendix. We suggest that the best evaluation of this position paper is to investigate the directions proposed and see whether the experiments lead to fruitful new avenues for interaction with LLMs.

\subsubsection*{Acknowledgments}
Diana Robinson thanks the Accelerate Programme for Scientific Discovery, made possible by a donation from Schmidt Sciences, for their funding support. Neil Lawrence is funded by grant EP/V030302/1 generously provided by the Engineering and Physical Sciences Research Council (EPSRC) and the Alan Turing Institute.

\bibliography{LLMs-visual-perception}
\bibliographystyle{iclr2025_conference}

\appendix
\section{Appendix}

\begin{table}[ht]
\small
\caption{Summary of key theories of visual perception and connections to possible research directions for LLMs.}
\label{summary-table}
\begin{center}
\begin{tabular}{p{0.19\linewidth}|p{0.37\linewidth}|p{0.37\linewidth}}
\toprule
\textbf{Research Theme} & \textbf{Theories of Visual Perception} & \textbf{Directions for LLM Research} \\
\midrule
Shared \\ Representations & External visual memory & Development of shared representations - e.g. mathematical proof as a shared representation in knowledge space \\
& & \\
 & Connection between physical exploration and visual exploration & Make connection between visual exploration and exploring the abstract information space \\
 & & \\
 & Ocular system mediating between the visual world and our perception of it & LLM mediating between our thinking and digital representations \\
 & & \\
 & Salience mapping—understanding where the eye fixates to develop an understanding of what information eyes rely on to process an image & Equivalent salience map of the information fixations of LLMs - can use this to develop understanding of depth and breadth of coverage across a set of knowledge \\
\midrule
Exploration and \\information seeking & Organisational model of information processing in the visual system & Relationship to how we filter information and how LLMs might do this with different emphasis \\
& & \\
 & Top-down, bottom-up and systematic saccade patterns & Using this as a model of understanding patterns of information exploration by both LLMs and their users and comparing the fit between these \\
 & & \\
 & Heavy-tailed distribution of saccadic eye movements & Using this as a guide to investigate whether LLMs explore the information space under a similar model \\
 & & \\
 & "Grabiness" of visual stimuli such as sudden noises or flashes of light & Develop similar mechanism in LLMs to draw attention to distribution shift and out of distribution data, signalling to the user to add in context \\
 \midrule
 Affordances & Affordances in visual world - based on information within them and our perceptual system - information that allows an animal to understand action possibilities in the environment & Provide a way of conceptualising and measuring domain understanding against representations of information provided by LLMs \\
\bottomrule
\end{tabular}

\end{center}
\end{table}

\end{document}